\documentclass[aps,prb,twocolumn]{revtex4}
\usepackage{graphicx}
\begin{document}
\title{Electronic structure of normal and inverse spinel ferrites from first principles}

\author{Z. Szotek$^{1}$, W.M. Temmerman$^{1}$, D. K\"{o}dderitzsch$^{2}$,
A. Svane$^{3}$, L. Petit$^{4}$, and H. Winter$^{5}$}
\affiliation{$^{1}$ Daresbury Laboratory, Daresbury, Warrington, WA4 4AD, Cheshire, U.K. \\
$^{2}$ Physikalische Chemie, Ludwig-Maximilian University, Munich, Germany \\
$^{3}$ Department of Physics and Astronomy, University of Aarhus, DK-8000 Aarhus C, Denmark \\
$^{4}$ Computer Science and Mathematics Division, and Center for Computational 
Sciences, Oak Ridge National Laboratory, Oak Ridge, TN 37831, USA \\
$^{5}$ IFP, Forschungszentrum Karlsruhe GmbH, Postfach 3640, D-76021 Karlsruhe, Germany}

\date{\today}

\begin{abstract}
We apply the self-interaction corrected local spin density 
approximation to study the electronic structure and magnetic properties of
the spinel ferrites MnFe$_{2}$O$_{4}$, Fe$_{3}$O$_{4}$, CoFe$_{2}$O$_{4}$, and 
NiFe$_{2}$O$_{4}$. We concentrate on establishing the nominal valence of 
the transition metal elements and the ground state structure, based on the
study of various valence scenarios for both the inverse and normal spinel 
structures for all the systems. 
For both structures we find all the studied compounds to be insulating, but 
with smaller gaps in the normal spinel scenario. On the contrary, the calculated 
spin magnetic moments and the exchange splitting of the conduction bands are seen
to increase dramatically when moving from the inverse spinel structure to
the normal spinel kind. We find substantial orbital moments for NiFe$_{2}$O$_{4}$ 
and CoFe$_{2}$O$_{4}$.
\end{abstract}

\maketitle

\section{Introduction}

The field of spintronics is concerned with search for highly spin-polarized
materials. One aim is to enhance tunnelling magnetoresistance (TMR) of magnetic
tunnel junctions (MTJs) which are active members of magnetic random access memory 
(MRAM) elements. Also, the highly spin-polarized materials are of paramount 
importance for increasing spin-polarization of 
currents injected into semiconductors, required for an optimal operation of spintronics 
devices.\cite{zutic} There appears to be a number of ways to achieve high spin-polarization, 
most notably by employing fully spin-polarized ferromagnetic metals, namely half-metals 
(HM).~\cite{coey} Another possibility is to exploit features of the band structure of such
tunnel barrier materials as e.g. MgO, and filtering electronic wave functions according 
to their symmetry in order to select the most highly spin-polarized ones.\cite{Parkin,Yuasa}
The least explored possibility is exploiting the spin-filtering effect, based on
ferromagnetic or ferrimagnetic insulating barriers. It was introduced by Moodera
et al.\cite{Moodera}, using EuS tunnel barriers. Spin filtering effect has been 
demonstrated in Gd/EuS/Al junctions \cite{LeClair}, which exhibit high magnetoresistance, 
but show no great prospects for technological applications, on account of the low Curie
temperature, T$_{c}$, of EuS. 

Spinel ferrites~\cite{brabers} have been studied for many years both regarding their 
magnetic behaviour and correlated nature in conjunction with their structural properties to 
increase their performance in high-frequency devices. Some of them
can probably be used as spin filters. 
Spin-dependent gap should result in spin-dependent barrier
for tunnelling of electrons through the insulator, giving rise to spin filtering.
Since the tunnelling probability depends exponentially on the barrier height, the
spin filtering efficiency can be very high. 
Candidates for spin filters include such spinel ferrites as
NiFe$_{2}$O$_{4}$, CoFe$_{2}$O$_{4}$ and MnFe$_{2}$O$_{4}$~\cite{Penicaud}. 
In particular, recently a spin filtering efficiency of up to 22\% by the NiFe$_{2}$O$_{4}$ barrier has
been reported by L\"{u}ders et al.~\cite{ulrike,agnes}
In addition, L\"{u}ders et al.~\cite{ulu,Manuel} 
have demonstrated TMR of 120\% in La$_{2/3}$Sr$_{1/3}$MnO$_{3}$/SrTiO$_{3}$/NiFe$_{2}$O$_{4}$ 
junctions, which corresponds to 45\% spin polarization for the conductive NiFe$_{2}$O$_{4}$
film, which stays constant up to about 300 K.

These spinel ferrites belong to the same family as magnetite (Fe$_{3}$O$_{4}$) which has been
most thoroughly studied both for its HM character and the famous charge order~\cite{Verwey,Wright}.
Many theoretical studies have been dedicated to magnetite, and in particular its charge
order (which will not be discussed in the present paper) using various approximations to density 
functional theory (DFT) such as local spin density
(LSD) approximation or generalized gradient approximation (GGA), as well as going beyond,
for example by invoking the Hubbard U through the LDA+U (local density approximation + U) approach,
or using the self-interaction corrected (SIC)-LSD method.~\cite{anisimov,antonov,leonov,guoFFO,madsen,Szotek1} 
For the other spinel ferrites most theoretical studies have been done with LSD,
GGA~\cite{Penicaud,Singh,guo1,guo2} or hybrid density functionals.\cite{Zuo} The former two approaches 
usually describe these materials to be 
half-metallic and not insulating, if no distortions are included. The reason is that 
the transition metal (TM) $d$ electrons in oxides (as well as $f$ electrons in rare earth
compounds)
are strongly correlated and cannot be adequately described within the standard band theory 
framework with such approximations as LSD or GGA, placing them too high in energy around
the Fermi level.
The SIC-LSD method~\cite{wmt}, on the other hand, provides better 
description of correlations than LSD, and was successfully applied to a variety of $d-$ and 
$f-$electron materials \cite{ASOG,ZS,Nature}.
In this paper we apply the self-interaction corrected local spin density (SIC-LSD)
approximation to study the electronic structure of spinel transition metal oxides
MnFe$_{2}$O$_{4}$, Fe$_{3}$O$_{4}$, CoFe$_{2}$O$_{4}$, and NiFe$_{2}$O$_{4}$. 
We concentrate on the nominal valence of the TM elements and electronic and magnetic properties 
of these systems in both normal and inverse spinel structures. The reason being that in most
cases these materials appear to exist as some mixture of those structures.

The paper is organized as follows.  An overview of the basic features of the SIC-LSD 
formalism is presented in the next section. Section III gives some computational details, 
whilst the results of the application of the SIC-LSD method to the spinel ferrites 
TMFe$_{2}$O$_{4}$ (with TM=Mn, Fe, Co and Ni), are presented and discussed in section IV. 
The paper is concluded in section V.

\section{Theory}

The basis of the SIC-LSD formalism is a self-interaction free total energy functional,
\( E^{SIC} \), obtained by subtracting from the LSD total energy functional,
\( E^{LSD} \), a spurious self-interaction of each occupied electron state
\( \psi _{\alpha } \)\cite{PZ81}, namely
\begin{equation}
\label{eq1}
E^{SIC}=E^{LSD}-\sum _{\alpha }^{occ.}\delta _{\alpha }^{SIC}.
\end{equation}
 Here \( \alpha  \) numbers the occupied states and the self-interaction correction
for the state \( \alpha  \) is
\begin{equation}
\delta _{\alpha }^{SIC}=U[n_{\alpha }]+E_{xc}^{LSD}[\bar{n}_{\alpha }],
\end{equation}
with \( U[n_{\alpha }] \) being the Hartree energy and \( E_{xc}^{LSD}[\bar{n}_{\alpha }] \)
the LSD exchange-correlation energy for the corresponding charge density \( n_{\alpha } \)
and spin density \( \bar{n}_{\alpha } \). It is the LSD approximation to the exact
exchange-correlation energy functional which gives rise to the spurious self-interaction.
The exact exchange-correlation
energy \( E_{xc} \) has the property that for any single electron spin density,
\( \bar{n}_{\alpha } \), it cancels exactly the Hartree energy, thus \begin{eqnarray}
U[{n}_{\alpha }]+E_{xc}[\bar{n}_{\alpha }]=0.\label{can}
\end{eqnarray}
In the LSD approximation this cancellation does not take place, and for well localized
states the above sum can be substantially different than zero. For extended states in
periodic solids the self-interaction vanishes.

The SIC-LSD approach can be viewed as an extension of LSD
in the sense that the self-interaction correction is only finite for spatially
localized states, while for Bloch-like single-particle states \( E^{SIC} \)
is equal to \( E^{LSD} \). Thus, the LSD minimum is also a local minimum of
\( E^{SIC} \). A question now arises, whether there exist other competitive
minima, corresponding to a finite number of localized states, which could benefit
from the self-interaction term without loosing too much
of the energy associated with band formation.
This is often the case for rather well localized electrons like the 3$d$
electrons in transition metal oxides or the 4\( f \) electrons in rare earth
compounds. It follows from minimization of Eq. (\ref{eq1}) that within the SIC-LSD
approach such localized electrons move in a different potential than the delocalized
valence electrons which respond to the effective LSD potential. 
Thus, by including
an explicit energy contribution for an electron to localize, the {\it ab initio} SIC-LSD
describes both localized and delocalized electrons on an equal footing, leading
to a greatly improved description of correlation effects over
the LSD approximation, as well as, to determination of valence. 


In order to make the connection between valence and localization more explicit,
it is useful to define the nominal valence \cite{Nature} as
\[
N_{val}=Z-N_{core}-N_{SIC},
\]
where $Z$ is the atomic number (26 for Fe), $N_{core}$ is the number of core
(and semi-core) electrons (18 for Fe), and $N_{SIC}$ is the number of localized,
i.e., self-interaction corrected, states (either five or six, respectively for
Fe$^{3+}$ and Fe$^{2+}$). Thus, in this formulation the valence is equal to the
integer number of electrons available for band formation. The localized electrons
do not participate in bonding. To find the nominal valence we
assume various atomic configurations, consisting of different numbers of localized
states, and minimize the SIC-LSD energy functional of Eq. (\ref{eq1}) with respect
to the number of localized electrons.  

The SIC-LSD formalism is governed by the
energetics due to the fact that for each orbital the SIC differentiates between
the energy gain due to hybridization of the orbital with the valence bands and the
energy gain upon localization of the orbital. Whichever wins determines if the
orbital is part of the valence band or not, and in this manner
also leads to the evaluation of the valence of elements involved.
The SIC depends on the choice of orbitals and its value
can differ substantially as a result of this. Therefore, one
has to be guided by the energetics in defining the most optimally
localized orbitals to determine the absolute energy minimum of
the SIC-LSD energy functional. The advantage of the SIC-LSD formalism is that
for such systems as transition metal oxides or rare earth compounds the lowest
energy solution will describe the situation where some single-electron states
may not be of Bloch-like form. Specifically, in oxides, Mn-, Co-,  Ni-, and 
Fe-3$d$ states may be assumed to be localized, but not the O 2$p$ states, because 
treating them as localized is energetically unfavourable. 

The SIC-LSD approach has been implemented \cite{wmt}
within the linear muffin-tin-orbital (LMTO) atomic sphere approximation (ASA)
band structure method~\cite{oka75}, in the tight-binding representation~\cite{AJ84}. 
In this method the polyhedral Wigner Seitz cell is approximated by slightly 
overlapping atom centered spheres, with a total volume equal to the actual crystal 
volume, while the electron wave functions are expanded in terms of the screened muffin-tin 
orbitals, and the minimization of \( E^{SIC} \) becomes non-linear in the expansion 
coefficients.  The so-called combined correction term~\cite{skriver} has been implemented 
and consistently applied to improve on the ASA. \\

\section{Computational Details}

The spinel ferrites of interest to the present study have a general chemical formula
of the form $AB$$_{2}$$O$$_{4}$ and crystallize in the face-centred cubic structure. In the
normal spinel structure $A$ is a divalent element atom, occupying tetrahedral A sites,
while $B$ is a trivalent element, sitting on the octahedral B sites.   
When $A$ is a trivalent element, and $B$ consists of equal numbers of divalent and
trivalent elements, distributed over crystalographically equivalent B1 and B2 octahedral
sites, then the spinel structure is referred to as the inverse kind.
In TM ferrites substantial off-stoichiometry and intersite disorder are often present
in samples, but are not considered in this paper. The high temperature phase of magnetite 
is known to have the inverse 
spinel structure, where $A$ atoms are Fe$^{3+}$ ions, and B sites are equally populated by 
Fe$^{2+}$ and Fe$^{3+}$ ions. Similarly, the NiFe$_{2}$O$_{4}$ system has been established experimentally 
to crystallize in the inverse spinel structure, with the A sites being Fe$^{3+}$ ions,
while B sites equally populated by Ni$^{2+}$ and Fe$^{3+}$ ions. The MnFe$_{2}$O$_{4}$, on the other 
hand, is considered to be predominantly of the normal spinel kind, as about 80\% of A sites 
are populated by Mn$^{2+}$ ions.~\cite{hastings,brabers} The CoFe$_{2}$O$_{4}$ material is considered to be
mostly an inverse spinel compound with about 80\% of divalent Co ions occupying octahedral 
sites.~\cite{brabers,sawatzky,braicovich} As the experimental situation with respect to the 
observed structures and TM valences is not fully established, and the computer simulation of 
the exact 
physical conditions is very difficult, in this paper we study both extremes, namely the normal 
and inverse spinel structures for all the systems. In addition, we investigate a number of different 
valence scenarios, defined in the following sections, to find the most energetically favourable 
solutions, be it only at zero temperature, for all the studied systems.

Regarding the magnetic structure of the ferrites, we assume that of magnetite, with 
the spins of the TM atoms on the tetrahedral sublattice being antiparallel to those of
the octahedral sublattice. Within a given sublattice the spins of all the TM atoms are 
arranged in parallel to one another.

\begin{table}[t!]
\caption{The lattice constants (a) and corresponding ASA radii (r$_{ASA}$) for the 
transition metal elements occupying tetrahedral ($tet$) and octahedral ($oct$) sites, 
and for oxygen ions (in atomic units) for all the studied spinel ferrites. The
radii of the empty spheres used in the calculations are not given, although typically
four to five different types were used, all of the order of 1.65 - 1.95 atomic units, 
depending on the system, however, never exceeding the size of the oxygen spheres.}
\begin{tabular}{cccccc}
\hline
System & a & r$_{ASA}$$^{tet}$ & r$_{ASA}$$^{oct}$ & r$_{ASA}$$^{O1}$ & r$_{ASA}$$^{O2}$ \\
\hline
MnFe$_{2}$O$_{4}$    & 16.08  &  2.204  &  2.588 &  1.979 &  1.979  \\
Fe$_{3}$O$_{4}$      & 15.87  &  2.329  &  2.752 &  1.848 &  1.848  \\
CoFe$_{2}$O$_{4}$    & 15.84  &  2.434  &  2.696 &  1.928 &  1.928  \\
NiFe$_{2}$O$_{4}$    & 15.78  &  2.425  &  2.686 &  1.920 &  1.920  \\
\hline
\end{tabular}
\label{latt}
\end{table}

The calculations have been performed for the experimental lattice parameters, 
whose values, together with the corresponding ASA radii, are given in Table \ref{latt}.~\cite{Penicaud}. 
For the basis functions, we have used $s$-, $p$-, and $d$-muffin-tin orbitals on all the transition 
metal atoms as the so-called low waves and on the oxygen the $s$- and $p$-orbitals have been
treated as low waves and the $d$-orbitals have been downfolded~\cite{Lambrecht}. For a better
space filling and to increase the number of basis functions, a set of empty spheres
has also been included in the calculations. For the empty spheres only the $s$ basis functions
have been treated as low waves, while both $p$- and $d$-orbitals have been downfolded.
All the calculations have been performed
in the scalar-relativistic mode, but for the calculated ground state configurations 
the spin-orbit coupling (SOC) was also included to calculate the orbital moment, 
in addition to the spin moment.

\section{Results and discussion}

\subsection{MnFe$_{2}$O$_{4}$}

As mentioned earlier, this compound is believed to be of predominantly normal spinel 
character. 
It is insulating, with a small gap of 0.04-0.06 eV as determined by transport 
experiments.~\cite{flores} Our calculations have adressed
the important issues of this system by realizing both normal ('N') and inverse ('I') spinel arrangements 
of ions on the tetrahedral
and octahedral sites. 
In addition, the all 3+ scenario, where all the sites are occupied exclusively
by the 3+ ions, has also been studied. Note that in the normal spinel environment the latter would
mean that Mn$^{3+}$ (four $d$ electrons are considered as localized) ions occupy the tetrahedral 
sites, while all the octahedral sites are exclusively
populated by Fe$^{3+}$ ions ('N3+' scenario). For the inverse spinel environment the tetrahedral ions 
would be of Fe$^{3+}$ type, with the B1 octahedral sites occuped by Mn$^{3+}$ ions and the B2 sites 
by Fe$^{3+}$ ions ('I3+' scenario). In Table \ref{table1}, we summarize the total energy differences 
for all the scenarios studied for MnFe$_{2}$O$_{4}$, in comparison with all the other spinel ferrites.

\begin{table}[h!]
\caption{Total energy differences (in eV per formula unit), calculated within SIC-LSD,
between the ground state configuration and other 
valence/structure scenarios for all studied spinel ferrites at the experimental lattice constant. 
The row marked by 'N' means normal spinel arrangement, where the tetrahedral sites are occupied by 
the divalent ions, namely Mn$^{2+}$ in MnFe$_{2}$O$_{4}$ compound, Fe$^{2+}$ in Fe$_{3}$O$_{4}$, 
Co$^{2+}$ in CoFe$_{2}$O$_{4}$ and finally Ni$^{2+}$ in NiFe$_{2}$O$_{4}$, while the octahedral 
sites are populated exclusively by Fe$^{3+}$ ions. Similarly, the row marked by 'I' means that the 
B1 sites are occupied by Mn$^{2+}$ in MnFe$_{2}$O$_{4}$, Fe$^{2+}$ in Fe$_{3}$O$_{4}$, 
Co$^{2+}$ in CoFe$_{2}$O$_{4}$ and Ni$^{2+}$ in NiFe$_{2}$O$_{4}$, with all the tetrahedral and 
B2 octahedral sites taken by the Fe$^{3+}$ ions. The notation 'I3+' means that tetrahedral sites 
and B2 octahedral sites in all the compounds are occupied by Fe$^{3+}$ ions, while the B1 octahedral 
sites are populated by Mn$^{3+}$ ions in MnFe$_{2}$O$_{4}$, Fe$^{3+}$ ions in Fe$_{3}$O$_{4}$, 
Co$^{3+}$ ions in CoFe$_{2}$O$_{4}$, and Ni$^{3+}$ ions in NiFe$_{2}$O$_{4}$. In the 'N3+' scenario 
all the octahedral sites, in all the compounds studied, are occupied by the Fe$^{3+}$ ions, and the 
tetrahedral sites are taken by Mn$^{3+}$ ions in MnFe$_{2}$O$_{4}$, Fe$^{3+}$ ions in Fe$_{3}$O$_{4}$, 
Co$^{3+}$ ions in CoFe$_{2}$O$_{4}$, and Ni$^{3+}$ ions in NiFe$_{2}$O$_{4}$. Note, that in 
the Fe$_{3}$O$_{4}$ case the latter two scenarios, namely 'N3+' and 'I3+', are equivalent.}

\begin{tabular}{ccccc}
\hline
Scenario & MnFe$_{2}$O$_{4}$ & Fe$_{3}$O$_{4}$ & CoFe$_{2}$O$_{4}$ & NiFe$_{2}$O$_{4}$ \\
\hline
I    &  0.58  &  1.54  &  0.20 &  0.00 \\
I3+  &  0.92  &  0.00  &  0.00 &  0.52 \\
N    &  0.00  &  2.46  &  1.09 &  1.66 \\
N3+  &  0.28  &  0.00  &  0.46 &  1.57 \\
\hline
\end{tabular}
\label{table1}
\end{table}
 
We find the normal spinel arrangement of ions to be the calculated ground state for MnFe$_{2}$O$_{4}$
(Table \ref{table1}), in agreement with the experimental evidence for predominantly 
normal spinel character of this compound. The ground state solution is followed closely by 
the all 3+ scenario, realized in the normal spinel environment ('N3+'), which lies only 0.28 
eV higher in energy. The inverse spinel solution is 0.58 eV higher, while the 'I3+' scenario 
is the most unfavourable state for MnFe$_{2}$O$_{4}$. 

\begin{figure}[t!]
\includegraphics[scale=.30,angle=-90]{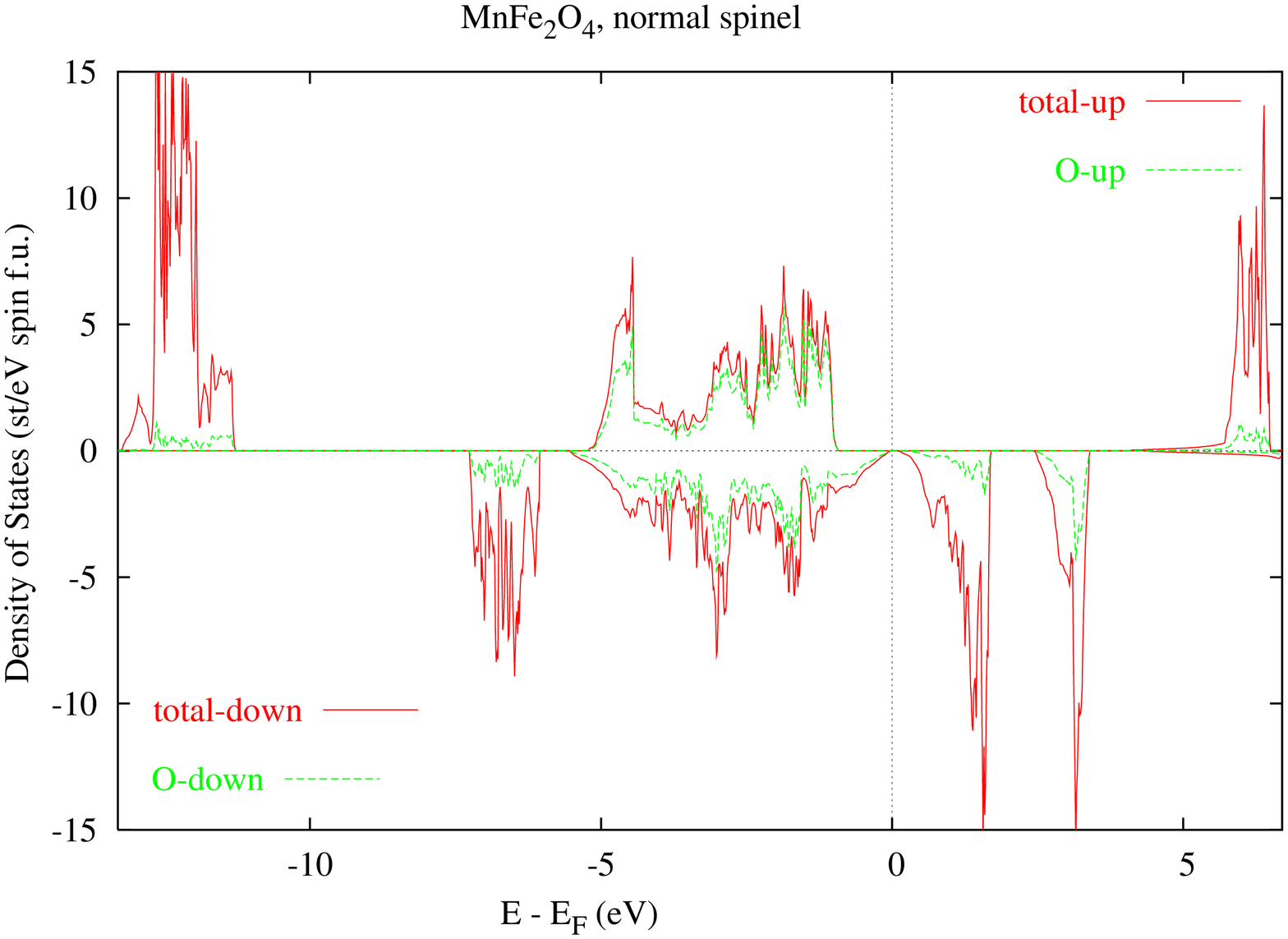} 
\includegraphics[scale=.30,angle=-90]{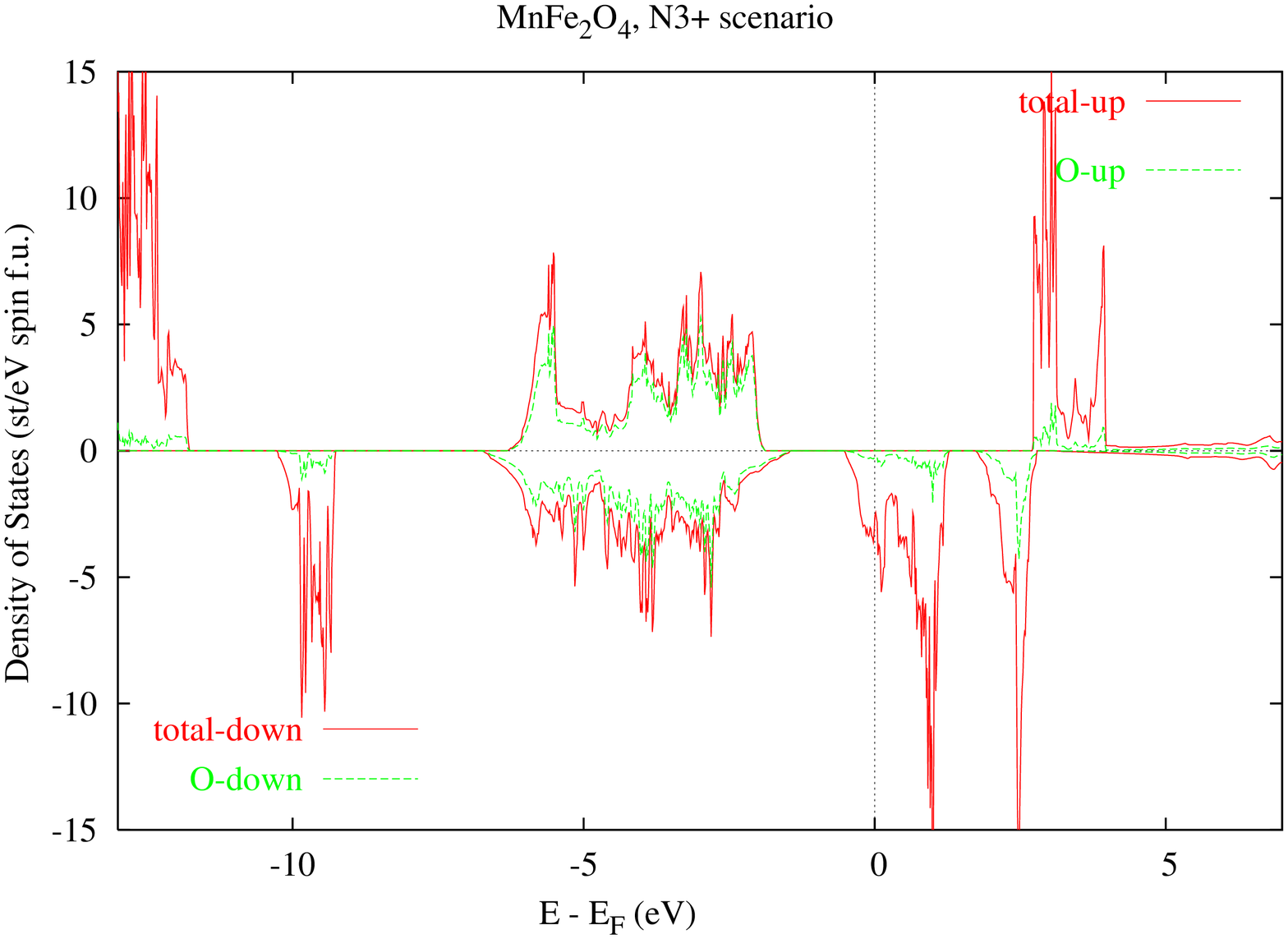}
\includegraphics[scale=.30,angle=-90]{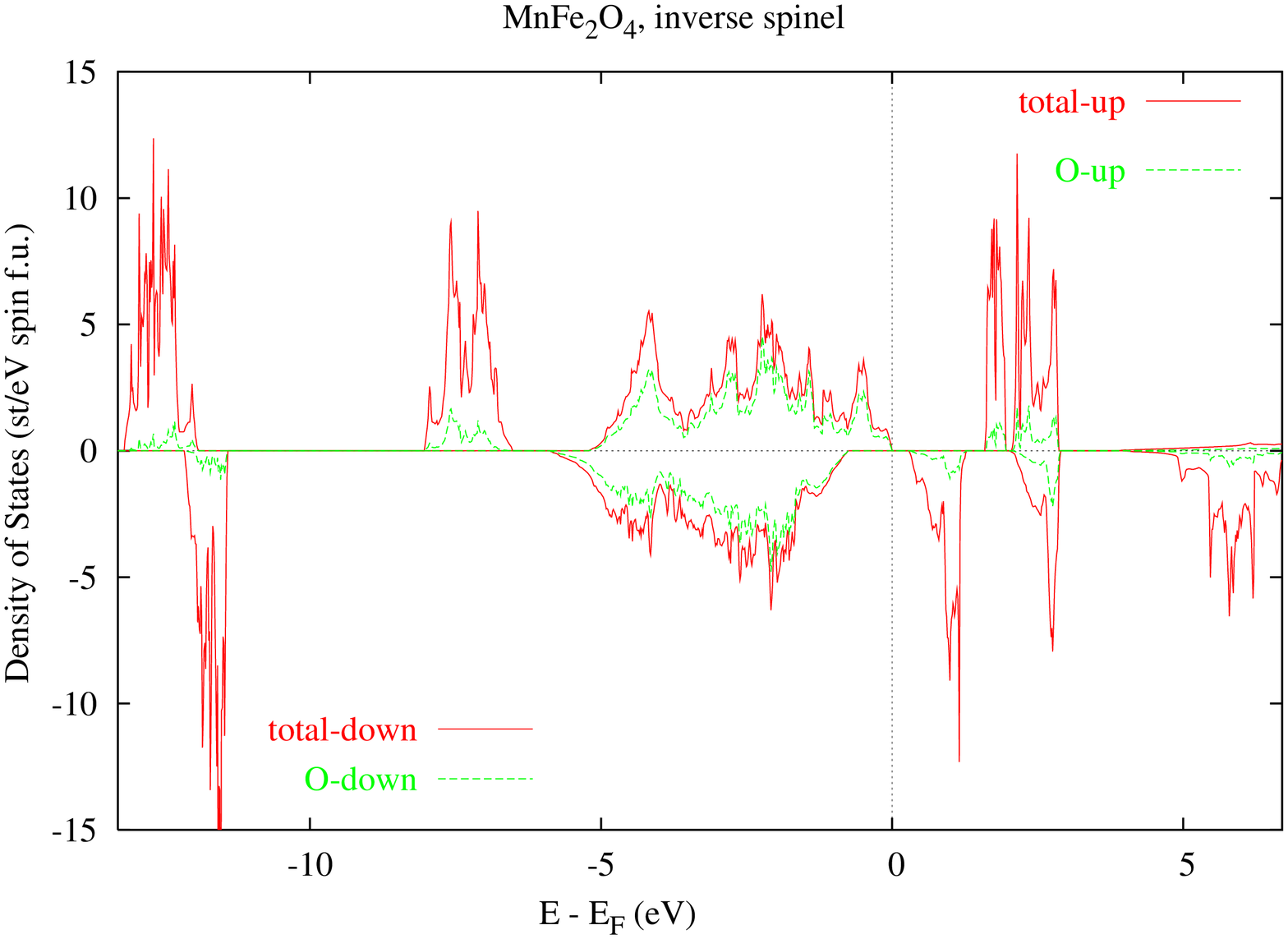}
\caption{(Color online) Spin decomposed total densities of states (in red), per formula unit, for 
MnFe$_{2}$O$_{4}$ for the normal spinel structure ('N') (top), 
for the all 3+ scenario in the normal spinel arrangement of atoms ('N3+') (middle), 
and for the inverse spinel scenario ('I') (bottom). 
The oxygen contribution to the total density of states is also shown (green, dotted
lines). 
The minority DOS is printed on the negative side of the y-axis, while the 
majority contributions are presented on the positive side of this axis.
\label{fig1}
}
\end{figure}

To establish whether the degree of localization of $d$ electrons of the TM ions residing
on the tetrahedral sites has any bearing on the preference towards normal spinel 
structure in MnFe$_{2}$O$_{4}$, we have looked at the change in the localization energy when switching
Mn$^{2+}$ ion between
the tetrahedral and octahedral sites. We have found that the localization energy
associated with Mn$^{2+}$ on the tetrahedral sites is 0.15 eV smaller than when Mn$^{2+}$
ions occupy the octahedral sites. At the same time the localization energy associated with
the Fe$^{3+}$ ions is smaller by 0.19 eV for the tetrahedral sites, in comparison with 
the situation when on the octahedral sites. Hence, the localization energy alone favours
the normal spinel structure only by 0.04 eV over the inverse spinel kind, which constitutes
just a tiny fraction of the total energy difference of 0.58 eV (Table \ref{table1}).
Thus, the preference of MnFe$_{2}$O$_{4}$ for the normal spinel structure is mostly driven by other
electronic degrees of freedom.

Regarding the density of states (DOS) of MnFe$_{2}$O$_{4}$ in the calculated ground state normal spinel 
structure (the top panel of Fig. \ref{fig1}), one 
can see that it is insulating, with a gap of about 0.075 eV which is slightly larger than
the experimental value obtained from transport measurements. A larger gap of about 0.3 eV is seen 
for the inverse 
spinel structure (bottom panel), but the spin splitting of the conduction band is here 
considerably smaller 
than for the ground state normal spinel scenario (Table \ref{tableMn}). In variance
to the normal and inverse spinel cases, the 'N3+' scenario (middle panel) is 
found to be half-metallic. 

For all the studied scenarios, the valence band is predominantly of the oxygen type, 
with a very small admixture of the TM character, and its polarization at the top of the band
changes from positive to negative between the inverse and normal spinel scenarios (Table \ref{tableMn}). 
What also changes substantially when moving from the normal to inverse spinel scenario is the
conduction band splitting, which for the normal spinel scenario is about 2.5 eV larger than that of
the inverse spinel case. The reason being that the unoccupied TM $d$-states are substantially pushed 
up in energy when in the normal spinel environment. Considering that in reality MnFe$_{2}$O$_{4}$ is not a pure
normal spinel compound, the exchange splitting of the conduction bands is most likely to be
somewhere in between the values calculated for the normal and inverse spinel scenarios. Of course,
the larger the splitting, the more advantageous should it be for the spin filtering properties.

\begin{table}
\caption{Spin decomposed exchange splittings of the valence and conduction bands, as well
as the energy gaps (in eV), for both inverse and normal spinel structures for MnFe$_{2}$O$_{4}$.
Here VBM stands for the valence band maximum and CBM for the conduction band minimum, and
$\uparrow$ refers to spin-up- and $\downarrow$ to spin-down-component. 
}
\begin{tabular}{ccc}
\hline
 & 'I' scenario & 'N' scenario \\
\hline
VBM$^{\uparrow}$ - VBM$^{\downarrow}$    &  0.69  & -0.91 \\
CBM$^{\uparrow}$ - CBM$^{\downarrow}$    &  1.31  &  3.85  \\
CBM$^{\uparrow}$ - VBM$^{\uparrow}$      &  1.64  &  4.84  \\
CBM$^{\downarrow}$ - VBM$^{\downarrow}$  &  1.02  &  0.075  \\
Gap                                      &  0.33  &  0.075  \\
\hline
\end{tabular}
\label{tableMn}
\end{table}
 
To understand details of the densities of states shown in Fig. \ref{fig1}, associated with
the localized TM states, one should keep in
mind that in the normal spinel scenario (top panel) which, as already mentioned, is the 
calculated ground state
structure for this compound, the tetrahedral sublattice is populated by Mn$^{2+}$
ions and that in this case five minority Mn $d$ electrons are described as localized states,
seen just below the predominantly O 2$p$ valence band, at about -6 eV. The unoccupied, majority
Mn $d$ electrons give rise to the states seen as peaks above the Fermi energy at about +6 eV.
The octahedral sublattice is populated by Fe$^{3+}$ ions, with five majority $d$ electrons 
localized, as the two sublattices are anti-parallel to one another. As a result the 
localized majority $d$ Fe states
are seen at about -12 eV below the Fermi energy. The unoccupied Fe $d$ states, the minority ones,
are seen just above the Fermi energy, over the range of up to about 3.5 eV. 

In the 'N3+' scenario (middle panel),
one minority Mn $d$ electron gets delocalized to realize Mn$^{3+}$ ions on the tetrahedral sites.
As a result the localized, minority Mn $d$ peak has moved down in energy, lying just above -10 eV,
while the fifth, now delocalized, minority $d$ electron appears at the Fermi energy, bringing 
down also the unoccupied, minority, Fe $d$ states. The situation changes in the inverse spinel
scenario (bottom panel), as now Mn$^{2+}$ ions reside on B1 sites of the octahedral sublattice, while 
the Fe$^{3+}$ ions populate the tetrahedral sublattice and B2 sites of the octahedral sublattice. 
Thus on the tetrahedral sites we have five minority Fe $d$ electrons which give rise to the peak
at about -12 eV, while the localized Fe $d$ electrons of the octahedral sublattice are seen
as the peak on the majority side, at about -14 eV. The five localized Mn majority $d$ states 
are again seen just below the valence band at about -7 eV. The unoccupied Mn minority $d$ bands
are seen at about 6 eV above the Fermi energy. The five unoccupied, majority, Fe $d$ states,
associated with the tetrahedral sublattice, are seen at about 2.5 eV above the Fermi energy.
Finally, the unoccupied, minority, Fe $d$ states of the octahedral sublattice are seen as
two separate peaks above the Fermi energy. Note, however, that the SIC-LSD eigenvalues
have no direct physical interpretations as removal energies, and thus should not be directly 
compared with spectroscopies.  To do so, one would need to take into account relaxation/screening 
effects that are not included in such an effective one-electron theory as SIC-LSD. One way to 
accomplish this is to employ the $\Delta_{SCF}$ calculations,\cite{DelSCF,WalterPr,photoemission_comment}
and another is the SIC-LSD based optimized effective potential (OEP) 
method.\cite{Sharp,Talman,OEP,photoemission_comment}
\begin{table}
\caption{Total spin magnetic moments (in $\mu_{B}$ per formula unit), calculated within SIC-LSD,
for all the studied spinel ferrites and scenarios.}

\begin{tabular}{ccccc}
\hline
Scenario & MnFe$_{2}$O$_{4}$ & Fe$_{3}$O$_{4}$ & CoFe$_{2}$O$_{4}$ & NiFe$_{2}$O$_{4}$ \\
\hline
I    &  5.00  &  4.00  &  3.00 &  2.00 \\
I3+  &  4.10  &  4.00  &  3.00 &  2.00 \\
N    &  5.00  &  6.00  &  7.00 &  8.00 \\
N3+  &  5.00  &  4.00  &  5.60 &  6.80 \\
\hline
\end{tabular}
\label{table2}
\end{table}

The magnetic properties change when moving from the normal to inverse spinel scenario, as seen
in Table \ref{table2}, where we compare the total spin magnetic moments for all the studied
spinel ferrites. The total spin magnetic moment for MnFe$_{2}$O$_{4}$ is 5 $\mu_{B}$ per formula unit, 
for both insulating and half-metallic solutions, while for the metallic 'I3+' scenario the
spin moment is reduced to 4.1 $\mu_{B}$ per formula unit. Note that unlike in the other ferrites
there is no change in the total spin magnetic moment between the normal and inverse spinel
scenarios. The reason being that, as seen in Table \ref{table3}, there are only very small changes
in the values of the spin moments of the transition metal ions, that are compensated by changes in
the induced oxygen spin moments.

\begin{table}[h!]
\caption{Type-decomposed spin magnetic moments (in $\mu_{B}$ per formula unit), 
calculated within SIC-LSD,
for MnFe$_{2}$O$_{4}$ for inverse and normal spinel structures. Here 'A' marks
the tetrahedral-, while 'B1' and 'B2' the octahedral-sites, and 'O1' and 'O2'
stand for two different types of oxygens.}

\begin{tabular}{cccccc}
\hline
Scenario & Fe$_{A}$$^{3+}$ & Mn$_{B1}$$^{2+}$ & Fe$_{B2}$$^{3+}$ & O1 & O2 \\
\hline
I    &  -4.09  &  4.58  &  4.11 &  0.12  & 0.03\\
\hline
Scenario & Mn$_{A}$$^{2+}$ & Fe$_{B1}$$^{3+}$ & Fe$_{B2}$$^{3+}$ & O1 & O2 \\
\hline
N    &  -4.49  &  4.11  &  4.11 &  0.34 & 0.34 \\
\hline
\end{tabular}
\label{table3}
\end{table}

Including spin-orbit coupling for the ground state normal spinel scenario for MnFe$_{2}$O$_{4}$ compound,
we find no considerable orbital moments either on Mn (-0.0005 $\mu_{B}$) or 
on Fe ions (0.019 $\mu_{B}$). 
As a result, the total orbital moment is of the order of 0.045 $\mu_{B}$ per formula 
unit, while at the same time the total spin moment is changed from 5.0 $\mu_{B}$ to
4.9995 $\mu_{B}$ per formula unit. Also, even with SOC included, we still observe 
a small energy gap of 0.064 eV, which incidently is in good agreement with the transport 
experiments.~\cite{flores}

\subsection{Fe$_{3}$O$_{4}$}

Based upon its high magnetoresistive properties, magnetite 
is of interest for technological applications, as e.g. computer
memory, magnetic recording, etc. 
Magnetite is
thought to be half-metallic, with the highest known T$_{c}$ of 860
K. At about T$_{V}$=122 K it undergoes a transition to
an insulating state, associated with some kind of charge
order, setting in on the octahedral sites, and a distortion of the crystal structure from
the inverse spinel cubic to monoclinic \cite{Verwey,Walz,Wright}.
Verwey argued that below the transition temperature, T$_{V}$, the Fe$^{3+}$ 
and Fe$^{2+}$ cations order in the alternate (001) planes, and interpreted this 
transition as an electron localization-delocalization transition \cite{Verwey}.

In the earlier paper, we have studied three different types of charge order on the octahedral 
sites, both in the high temperature (cubic) and low temperature (monoclinic) phases~\cite{Szotek1}. 
In this paper, for the sake of comparison with other spinel TM oxides, we concentrate exclusively 
on the high temperature cubic phase and the scenarios enumerated in Table \ref{table2}.

\begin{figure}[h]
\includegraphics[scale=.30,angle=-90]{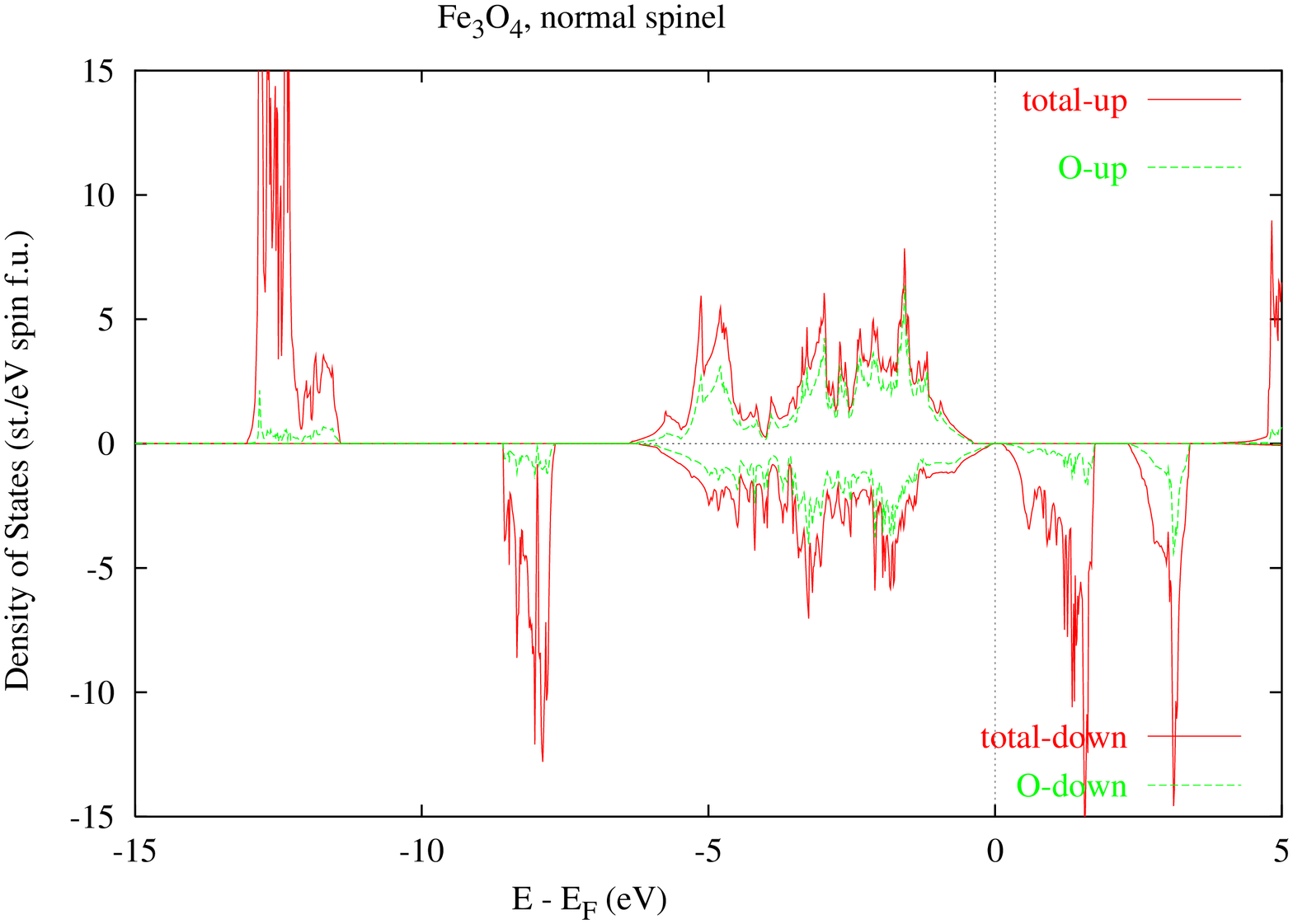} 
\includegraphics[scale=.30,angle=-90]{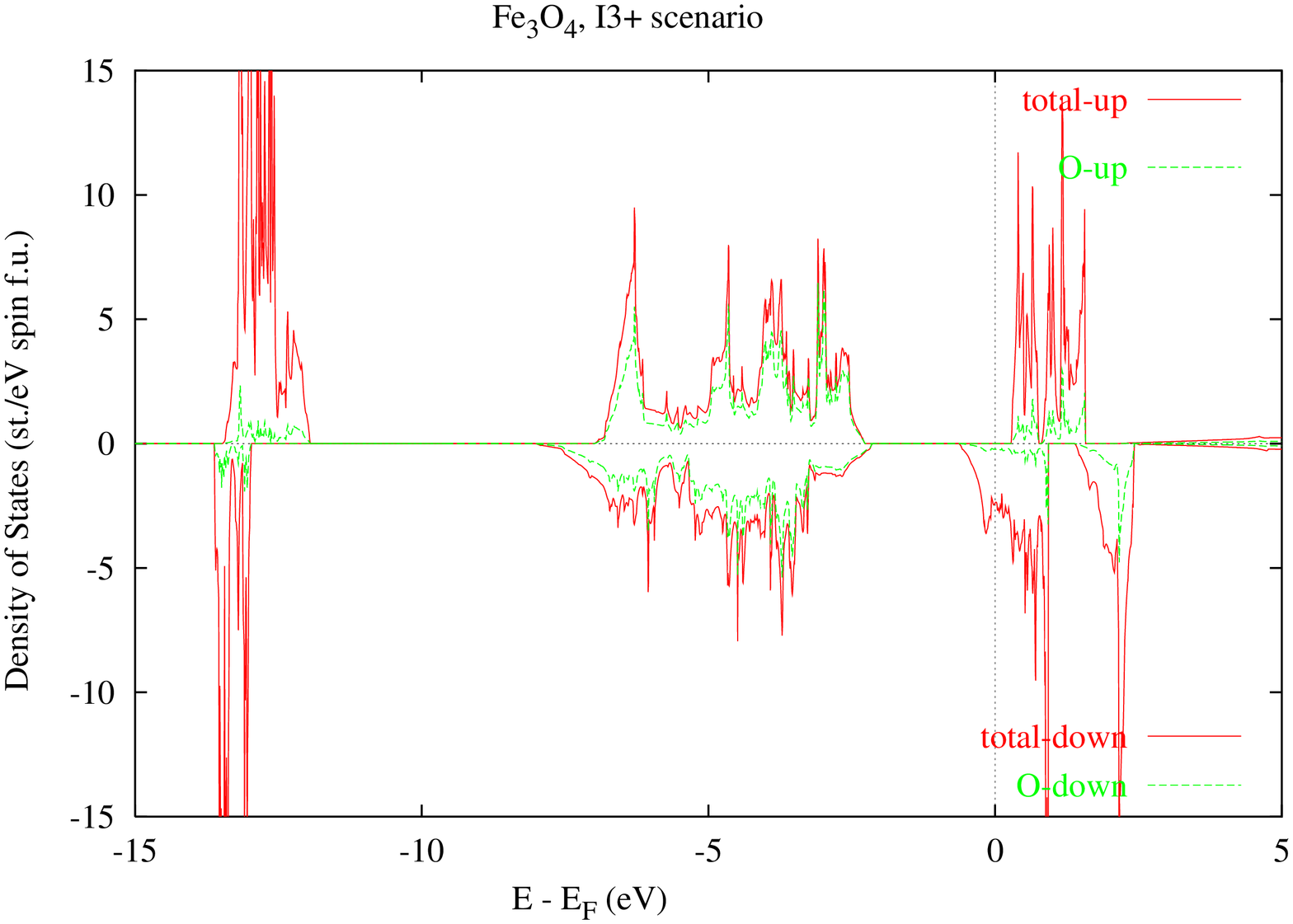}
\includegraphics[scale=.30,angle=-90]{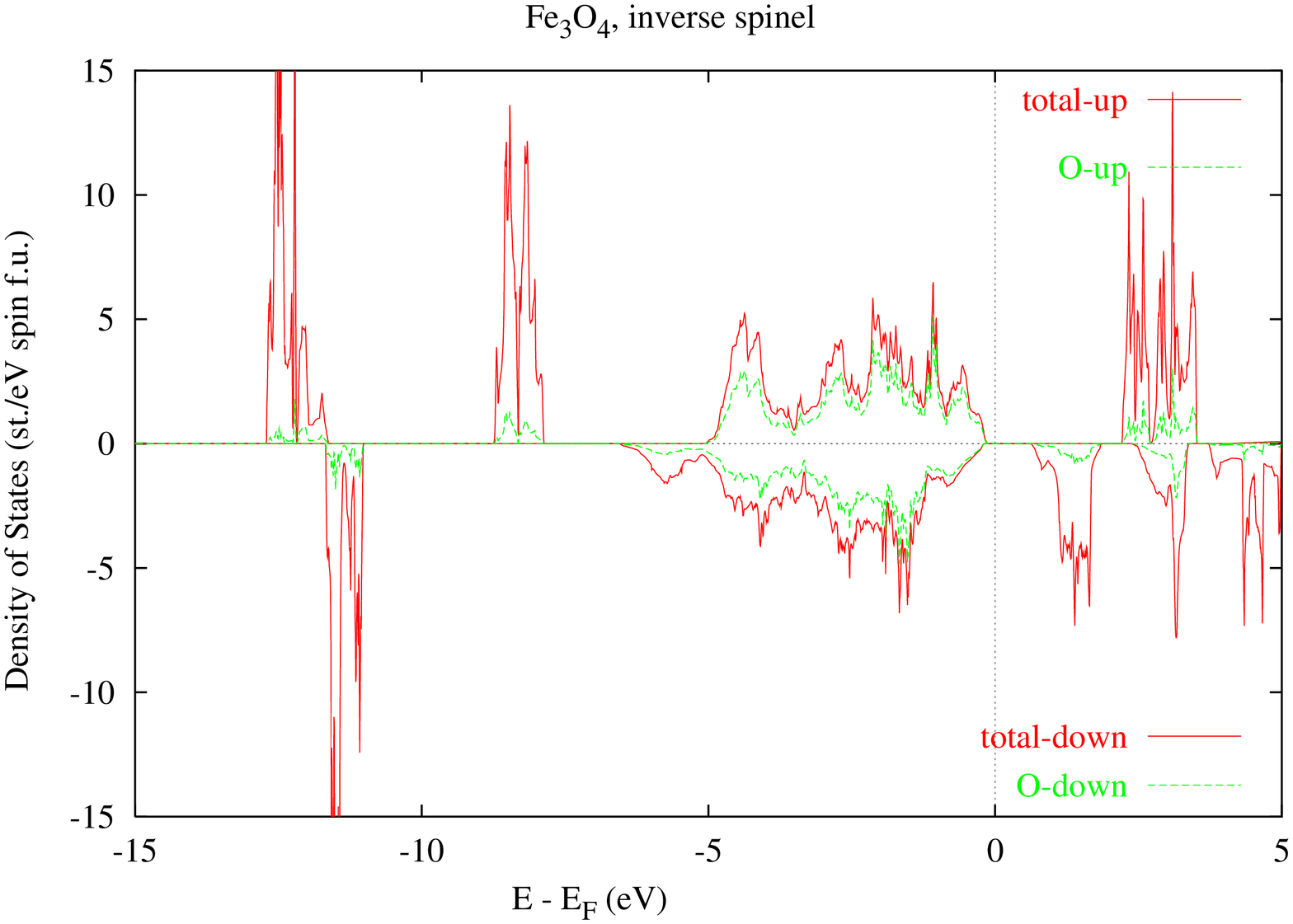}
\caption{(Color online) Spin decomposed total densities of states (in red), per formula unit, for 
Fe$_{3}$O$_{4}$ for the normal spinel structure ('N') (top), 
for the all 3+ scenario in the inverse spinel arrangement of atoms ('I3+') (middle), 
and for the inverse spinel scenario ('I') (bottom). 
The oxygen contribution to the total density of states is also shown (green dotted).  
The minority DOS is printed on the negative side of the y-axis, while the majority 
contribution is shown on the positive side of this axis.
\label{fig2}
}
\end{figure}

As seen from Table \ref{table1} and Fig. \ref{fig2} (middle panel), the ground state of 
magnetite in the cubic
phase is half-metallic, with all Fe ions in 3+ configuration (five $d$ electrons localized).
This ground state scenario ('I3+'$\equiv$'N3+') is the result of a delocalization of the
sixth $d$ electron of the original Fe$^{2+}$ ions, that together with Fe$^{3+}$ ions
randomly populate the octahedral B1 and B2 sites in the high-temperature cubic phase.
In the ground state scenario, this sixth electron is seen to give rise to the peak at 
the Fermi energy in the minority channel, together with the other 10 unoccupied minority 
$d$ states associated with the octahedral sites. 
The five localized tetrahedral Fe$_{A}$ minority $d$ states appear around -13 eV, while the 
unoccupied majority Fe$_{A}$ $d$ states are seen just above the Fermi energy. All the 
localized majority $d$ states of the octahedral sites are at about -12.5 eV.
The valence band is of predominantly O $p$ character.

The inverse spinel solution (Table \ref{table1} and the bottom panel of Fig. \ref{fig2}), 
corresponding to the assumed Verwey charge order, lies about 1.5 eV above the ground 
state. In this scenario, the above mentioned sixth electron is localized on the B1-octahedral 
sites, and appears as a small hump at the bottom of the minority valence band. The remaining 
five $d$ states of the Fe$_{B1}$ sites are seen at about -8 eV below the Fermi energy in 
the majority bands. The five localized majority Fe$_{B2}$ $d$ states lie at about -12.5 eV.  
The minority Fe$_{A}$ $d$ states (seen below -10 eV) are localized, and the majority 
Fe$_{A}$ $d$ states are unoccupied, occuring at about 3 eV above the Fermi energy. 
As a result, since the four remaining minority Fe$_{B1}$ $d$ states and five
minority Fe$_{B2}$ states are also unoccupied, our calculations for this scenario
give an insulating state with a gap of $\sim$0.7 eV. Understandably, as the latter has been
calculated for the high-temperature inverse spinel structure, its value is much 
larger than the experimental value of 0.14 eV~\cite{gap}, measured for the true,
low temperature monoclinic phase.\cite{gap1}
%

Our calculations for the normal spinel structure give the most energetically
unfavourable solution for magnetite, lying about 2.5 eV above the 'I3+' ground state scenario.
In this normal spinel case (top panel of Fig. \ref{fig2}), we obtain an insulating
solution with an energy gap of 0.08 eV. Here, the tetrahedral sites are occupied by
Fe$^{2+}$ ions, while the octahedral sites are populated with Fe$^{3+}$ ions.
As a result, five localized minority Fe$_{A}$ $d$ states lie at about -8 eV,
while the sixth localized, majority, Fe$_{A}$ $d$ state sits right at the bottom
of the majority valence band. All the remaining unoccupied majority Fe$_{A}$ $d$ 
states are just about seen at 5 eV above the Fermi energy, giving rise to a large
exchange splitting of the conduction band. The localized, majority, Fe$_{B1}$ and  
Fe$_{B2}$ $d$ states are seen at -12.5 eV, while their unoccupied, minority,
states lie just above the Fermi energy, over the range of about 3-4 eV. So, again
like in MnFe$_{2}$O$_{4}$, we see a large change in the exchange splitting of the conduction band
when moving from the inverse to normal spinel structure.
For the inverse spinel scenario ('I') it is of the order of 1.6 eV, while for the
normal spinel arrangement it increases to 3.65 eV (Table \ref{tableFe}). 
It is interesting to note, that in the normal spinel structure the sixth localized
$d$ electron of the Fe$^{2+}$ ion occupies one of the e$_{g}$ orbitals, while in the
inverse spinel scenario it populates one of the t$_{2g}$ states.

\begin{table}
\caption{Spin decomposed exchange splittings of the valence and conduction bands, as well
as the energy gaps (in eV), for both inverse and normal spinel structures for Fe$_{3}$O$_{4}$.
Here VBM stands for the valence band maximum and CBM for the conduction band minimum, and
$\uparrow$ refers to spin-up- and $\downarrow$ to spin-down-component. 
}
\begin{tabular}{ccc}
\hline
 & 'I' scenario & 'N' scenario \\
\hline
VBM$^{\uparrow}$ - VBM$^{\downarrow}$    &  0.06  & -0.35 \\
CBM$^{\uparrow}$ - CBM$^{\downarrow}$    &  1.61  &  3.65  \\
CBM$^{\uparrow}$ - VBM$^{\uparrow}$      &  2.33  &  4.08  \\
CBM$^{\downarrow}$ - VBM$^{\downarrow}$  &  0.78  &  0.08  \\
Gap                                      &  0.72  &  0.08  \\
\hline
\end{tabular}
\label{tableFe}
\end{table}
 
\begin{table}
\caption{Total and type-decomposed spin magnetic moments (in Bohr magnetons per
formula unit) for magnetite as calculated within SIC-LSD for three different
scenarios. Only different Fe-types are listed in the table.
}
\begin{tabular}{ccccccc}
\hline
Scenario & M$_{total}$ & M$_{Fe_{A}^{2+}}$ & M$_{Fe_{A}^{3+}}$ & M$_{Fe_{B1}^{2+}}$ & M$_{Fe_{B1}^{3+}}$ & M$_{Fe_{B2}^{3+}}$  \\
\hline
I        & 4.00 &  -    & -4.00 & 3.57 & -     & 4.08  \\
I3+      & 4.00 &  -    & -4.02 & -    &  3.90 & 3.90  \\
N        & 6.00 & -3.46 &  -    & -    &  4.09 & 4.09  \\
\hline
\end{tabular}
\label{table4}
\end{table}

The total spin magnetic moment per formula unit is 4 $\mu_{B}$ (Table \ref{table2}) for all 
the scenarios studied, with the exception of the normal spinel scenario, where we see
a 50\% increase to 6 $\mu_{B}$. As all the scenarios give rise to either 
insulating or half-metallic states, the spin magnetic moments are naturally integer numbers. 
Table \ref{table4} explains how the 50\% increase in the total spin magnetic moment comes
about when switching from the inverse to normal spinel structure. In the inverse spinel case
the spin moment of the tetrahedral Fe ions gets just about cancelled by the spin moment
of the B2-octahedral sites, so that the total spin moment is mostly due to the Fe$^{2+}$
ions on the B1-sites. In the normal spinel, on the other hand, the spin moment of the
tetrahedral Fe$^{2+}$ ions is smaller, and oppositely alligned with the spin moments of
the Fe$^{3+}$ ions that occupy all the octahedral sites. Bearing in mind that there are twice as
many octahedral sites as the tetrahedral ones, the substantial increase is easy to account
for, especially that the induced spin moments on the oxygen sites do not differ much between
the two scenarios.

Including spin-orbit coupling for the ground state 'I3+' scenario leads
to a very small total orbital moment of about 0.05 $\mu_B$ per formula unit, while the total
spin moment is very slightly reduced from 4 $\mu_B$ to 3.9998 $\mu_B$ per formula unit. The 
orbital moments due to the individual Fe-ions are similarly very small, with the tetrahedral 
Fe being -0.015 $\mu_B$ and the octahedral Fe of 0.035 $\mu_B$.


\subsection{CoFe$_{2}$O$_{4}$}

\begin{figure}[t!]
\includegraphics[scale=.30,angle=-90]{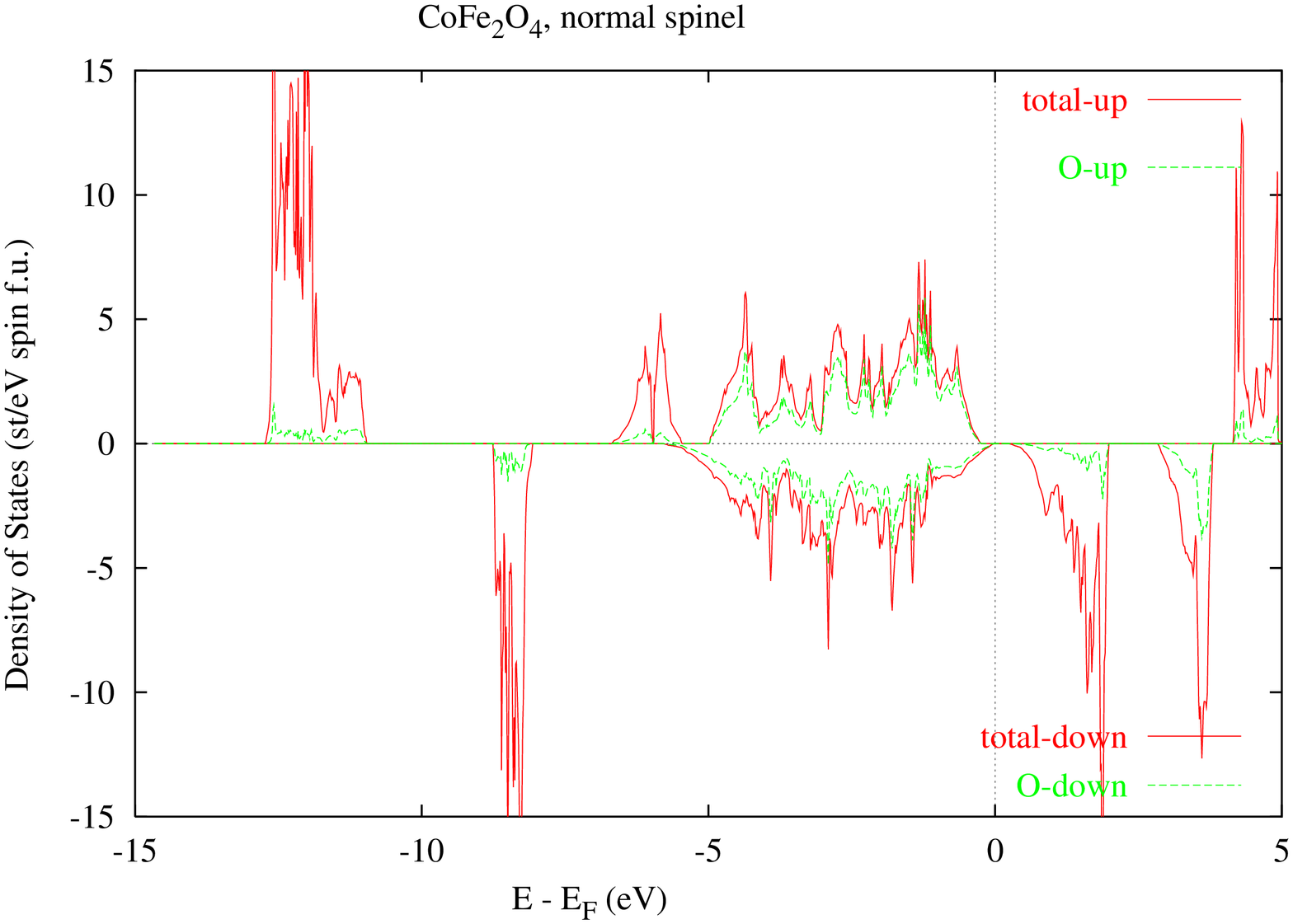} 
\includegraphics[scale=.30,angle=-90]{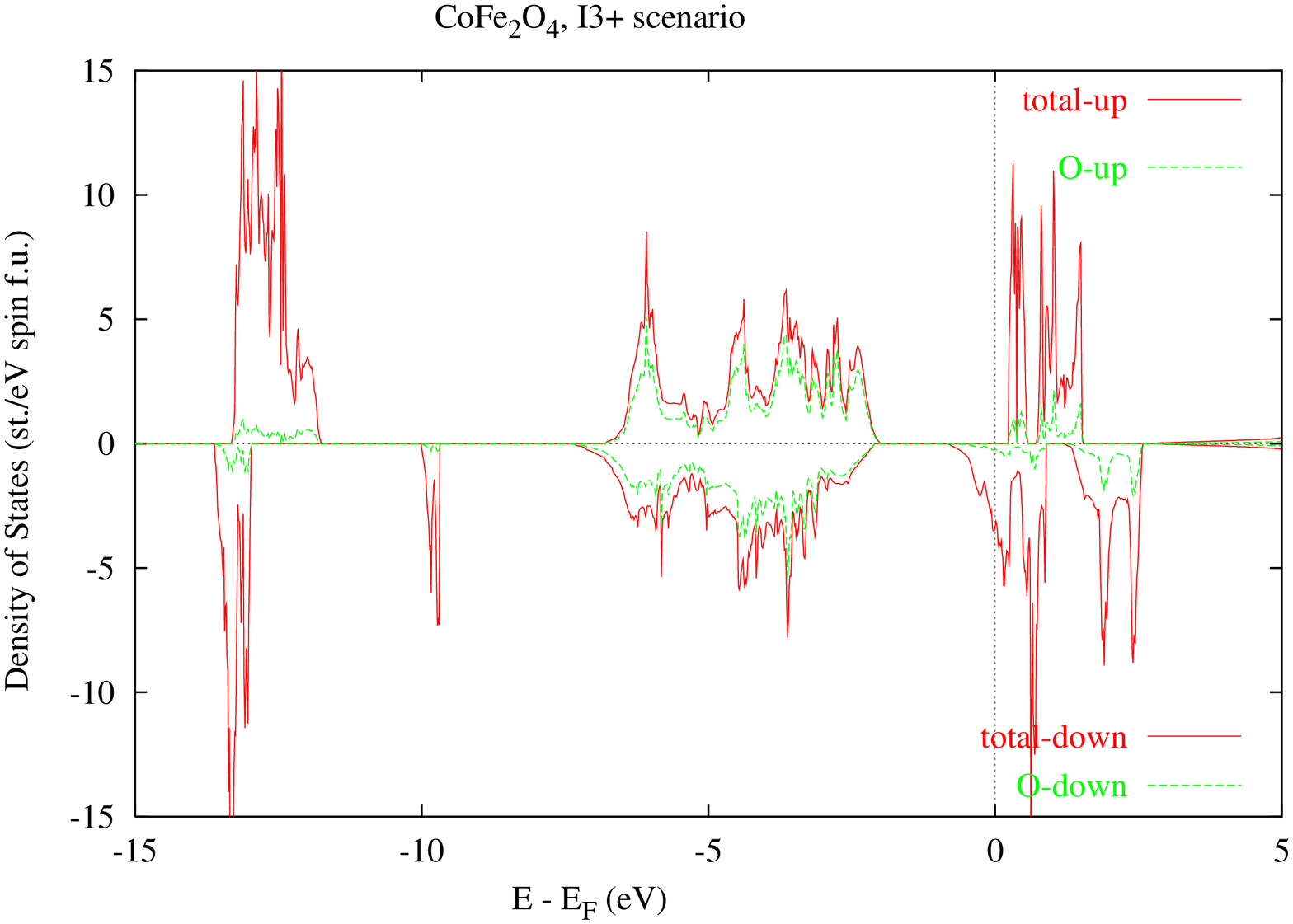}
\includegraphics[scale=.30,angle=-90]{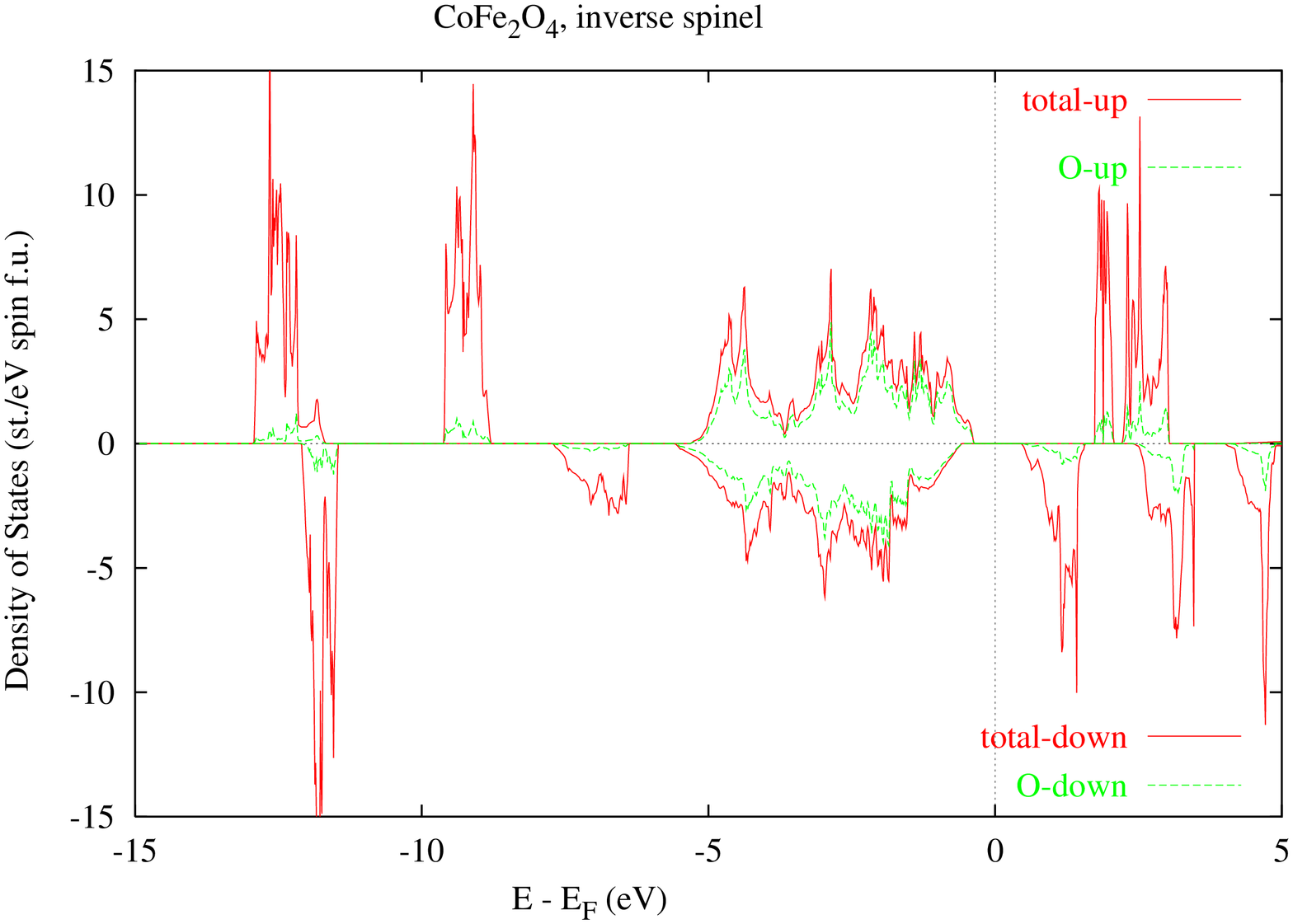}
\caption{(Color online) Spin decomposed total densities of states (in red), per formula unit, for 
CoFe$_{2}$O$_{4}$ for the normal spinel structure ('N') (top), 
for the all 3+ scenario in the inverse spinel arrangement of atoms ('I3+') (middle), 
and for the inverse spinel scenario ('I') (bottom). 
The oxygen contribution to the total density of states is also shown (green dotted
lines).  
The minority DOS is printed on the negative side of the y-axis, while the
majority contribution is shown on the positive side of this axis.
\label{fig3}
}
\end{figure}

This compound is believed to be mostly of the inverse spinel kind~\cite{pattrick,waseda,neill}, 
with divalent Co ions occupying predominantly the octahedral sites. However, similarly to magnetite 
in the high temperature cubic 
phase, our calculations find the ground state of CoFe$_{2}$O$_{4}$ to be half-metallic and of the 
'I3+' type (Table \ref{table1} and Fig. \ref{fig3}, middle panel). As seen in Table \ref{table1}, 
the inverse spinel scenario ('I') is not far, lying only 
0.2 eV higher in energy. What this seems to imply is that this compound prefers the inverse 
arrangement of atoms, independently of the actual valence of the Co ions. The normal spinel
solution is about 1.1 eV away. Although the ground state we find is half-metallic,
the inverse spinel scenario ('I'), with Co$^{2+}$ ions occupying the B1 
octahedral positions, describes CoFe$_{2}$O$_{4}$ as an insulator, with a gap of 0.8 eV, which is reduced
to 0.21 eV in the normal spinel case (Table \ref{tableCo} and Fig. \ref{fig3}).

To understand in detail the densities of states in Fig. \ref{fig3} note
that Co has only one minority electron more than Fe. So, the Co$^{2+}$
ion has two localized minority $d$ electrons, in addition to the five majority ones.
In the ground state 'I3+' scenario (middle panel in Fig. \ref{fig3}), one of these two minority
electrons gets delocalized, contributing to the states  at the Fermi energy, while the other, 
localized, one is seen as a sharp peak just above -10 eV. All the remaining main features 
of DOS for this scenario are exactly like in the case of 'I3+' scenario in magnetite.

For the inverse spinel structure (bottom panel of Fig. \ref{fig3}), the situation is again 
very much like in magnetite (bottom panel of Fig. \ref{fig2}), with the exception that
now we have a small double hump, at about -7 eV, slightly detached from the predominantly 
O 2$p$ valence band, while in magnetite it was still attached to the valence band and
represented just a single minority $d$ electron. 

In the normal spinel scenario the Co$^{2+}$ ions now
reside on the tetrahedral sites, with their five localized minority $d$ states seen as
a rather sharp peak at about -8 eV, in the top panel of Fig. \ref{fig3}. The 
remaining two localized majority $d$ states 
are sitting just below the majority valence band. All the other features are like in 
magnetite. 

Similarly to other ferrites, one sees substantial change in the exchange 
splitting of the conduction band between the inverse and normal spinel scenarios, from
1.28 eV to 4.07 eV. Also, the negative spin polarization of the valence band is seen in
the normal spinel, while a positive one in the inverse spinel structure. Like in 
magnetite, for the divalent Co$^{2+}$ ions on the tetrahedral sites in the normal
spinel structure, the e$_{g}$ minority states are populated before the t$_{2g}$ ones,
which is in variance to the inverse spinel structure. In the latter case, the Co$^{2+}$ ions
reside on the B1 octahedral sites, and the t$_{2g}$ states are lying lower in energy than 
the e$_{g}$ states.

\begin{table}
\caption{Spin decomposed exchange splittings of the valence and conduction bands, as well
as the energy gaps (in eV), for both inverse and normal spinel structures for CoFe$_{2}$O$_{4}$.
Here VBM stands for the valence band maximum and CBM for the conduction band minimum, and
$\uparrow$ refers to spin-up- and $\downarrow$ to spin-down-component. 
}
\begin{tabular}{ccc}
\hline
 & 'I' scenario & 'N' scenario \\
\hline
VBM$^{\uparrow}$ - VBM$^{\downarrow}$    &  0.22  & -0.24 \\
CBM$^{\uparrow}$ - CBM$^{\downarrow}$    &  1.28  &  4.07  \\
CBM$^{\uparrow}$ - VBM$^{\uparrow}$      &  2.08  &  4.52  \\
CBM$^{\downarrow}$ - VBM$^{\downarrow}$  &  1.02  &  0.21  \\
Gap                                      &  0.80  &  0.21  \\
\hline
\end{tabular}
\label{tableCo}
\end{table}
 
As seen in Table \ref{table2}, the total spin magnetic moment for both the 'I' and 'I3+' 
scenarios is 3 $\mu_{B}$ per formula unit. It is reduced from 4 $\mu_{B}$ in magnetite due
to the smaller value of the spin moment of the divalent Co-ion  of 2.58 $\mu_{B}$ 
(Table \ref{table5}), in comparison with the spin moment of the divalent Fe ions of
3.57 $\mu_{B}$ (Table \ref{table4}). What is however more dramatic is the change of
the spin moment when moving from the inverse to the normal spinel arrangement of ions. 
Table \ref{table2} shows that in the normal spinel scenario the total spin moment is 7 $\mu_{B}$
per formula unit, which again is due to the fact that the octahedral sites are populated 
exclusively by
Fe$^{3+}$ ions, whose spin moments are arranged in parallel to one another and whose
magnitudes are considerably larger than the moment of Co$^{2+}$ ions on the tetrahedral sites.

\begin{table}
\caption{Type-decomposed spin magnetic moments (in $\mu_{B}$ per formula unit), 
calculated within SIC-LSD,
for CoFe$_{2}$O$_{4}$ for inverse and normal spinel scenarios. Here 'A' marks
the tetrahedral-, while 'B1' and 'B2' the octahedral-sites, and 'O1' and 'O2'
stand for two different types of oxygens.}

\begin{tabular}{cccccc}
\hline
Scenario & Fe$_{A}$$^{3+}$ & Co$_{B1}$$^{2+}$ & Fe$_{B2}$$^{3+}$ & O1 & O2 \\
\hline
I    &  -4.11  &  2.58  &  4.11 &  0.13  & 0.07\\
\hline
Scenario & Co$_{A}$$^{2+}$ & Fe$_{B1}$$^{3+}$ & Fe$_{B2}$$^{3+}$ & O1 & O2 \\
\hline
N    &  -2.58  &  4.13  &  4.13 &  0.32 & 0.32 \\
\hline
\end{tabular}
\label{table5}
\end{table}

With respect to spin-orbit coupling we find the total orbital moment of the ground
state 'I3+' scenario to be quite substantial of the order of 0.58 $\mu_{B}$ per
formula unit, and associated mostly with the Co$^{3+}$ ion. As the total spin
moment is reduced from 3 $\mu_{B}$ to 2.9997 $\mu_{B}$ per formula unit, the
ratio of the total orbital to spin moment is 0.19. The ratio of the orbital to
spin moment for the Co$^{3+}$ ion itself is 0.21.

\subsection{NiFe$_{2}$O$_{4}$}

NiFe$_{2}$O$_{4}$ is a ferromagnetic insulator that is of possible interest as a spin 
filter in MTJs \cite{ulrike,agnes}. This compound has the Curie temperature of 850 K,
and hence has a great potential for technological applications.
%
%

\begin{figure}[t!]
\includegraphics[scale=.30,angle=-90]{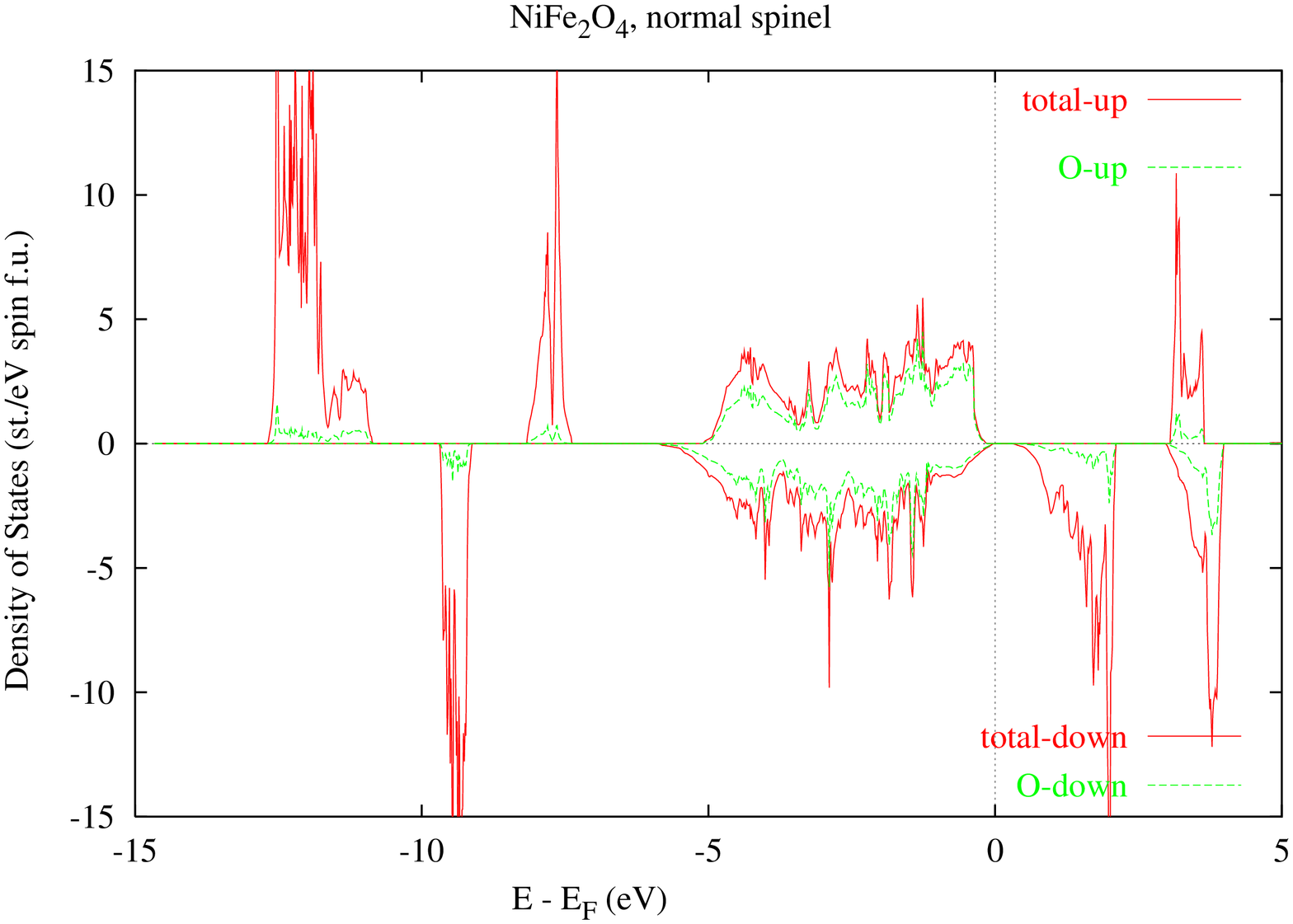} 
\includegraphics[scale=.30,angle=-90]{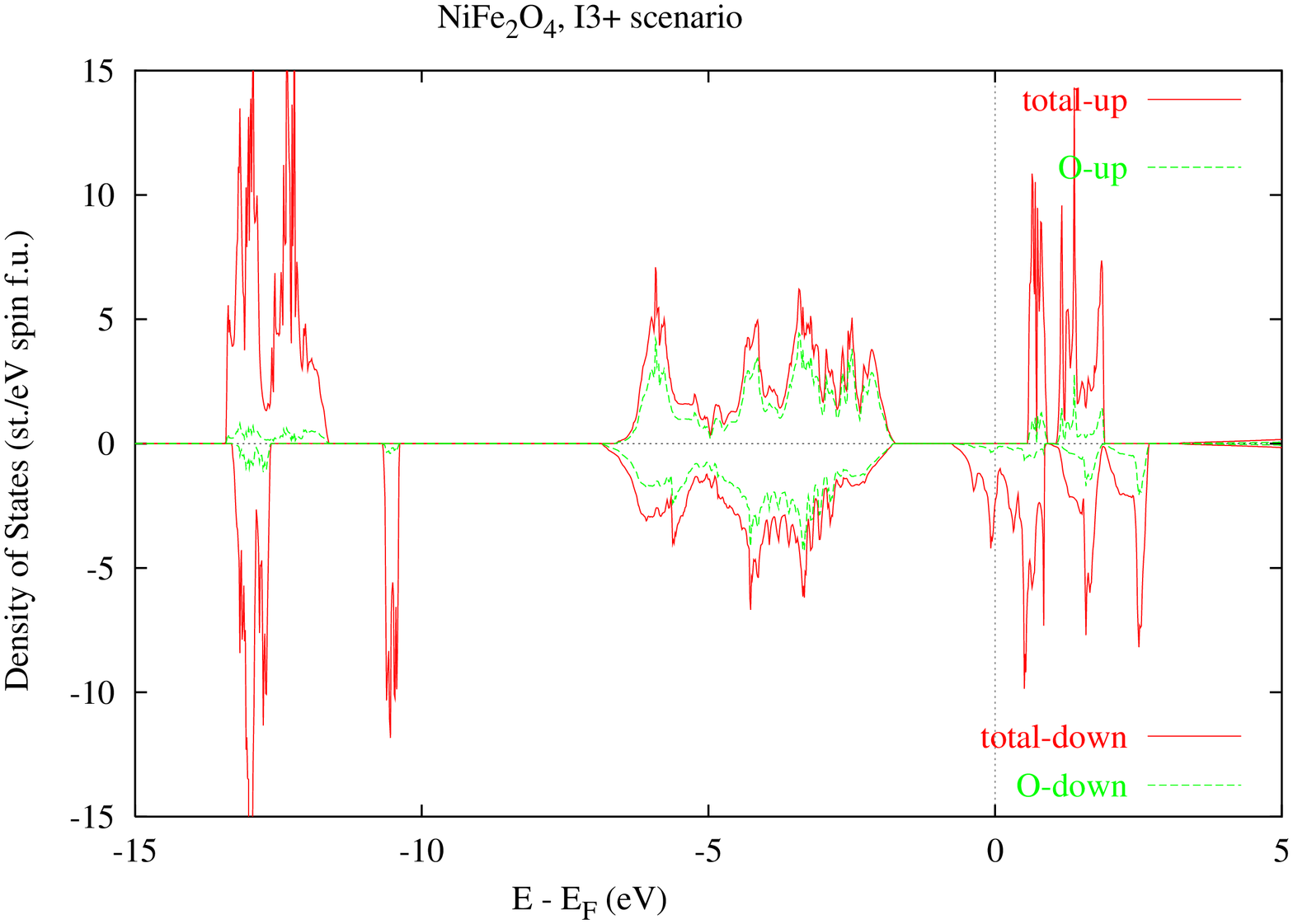}
\includegraphics[scale=.30,angle=-90]{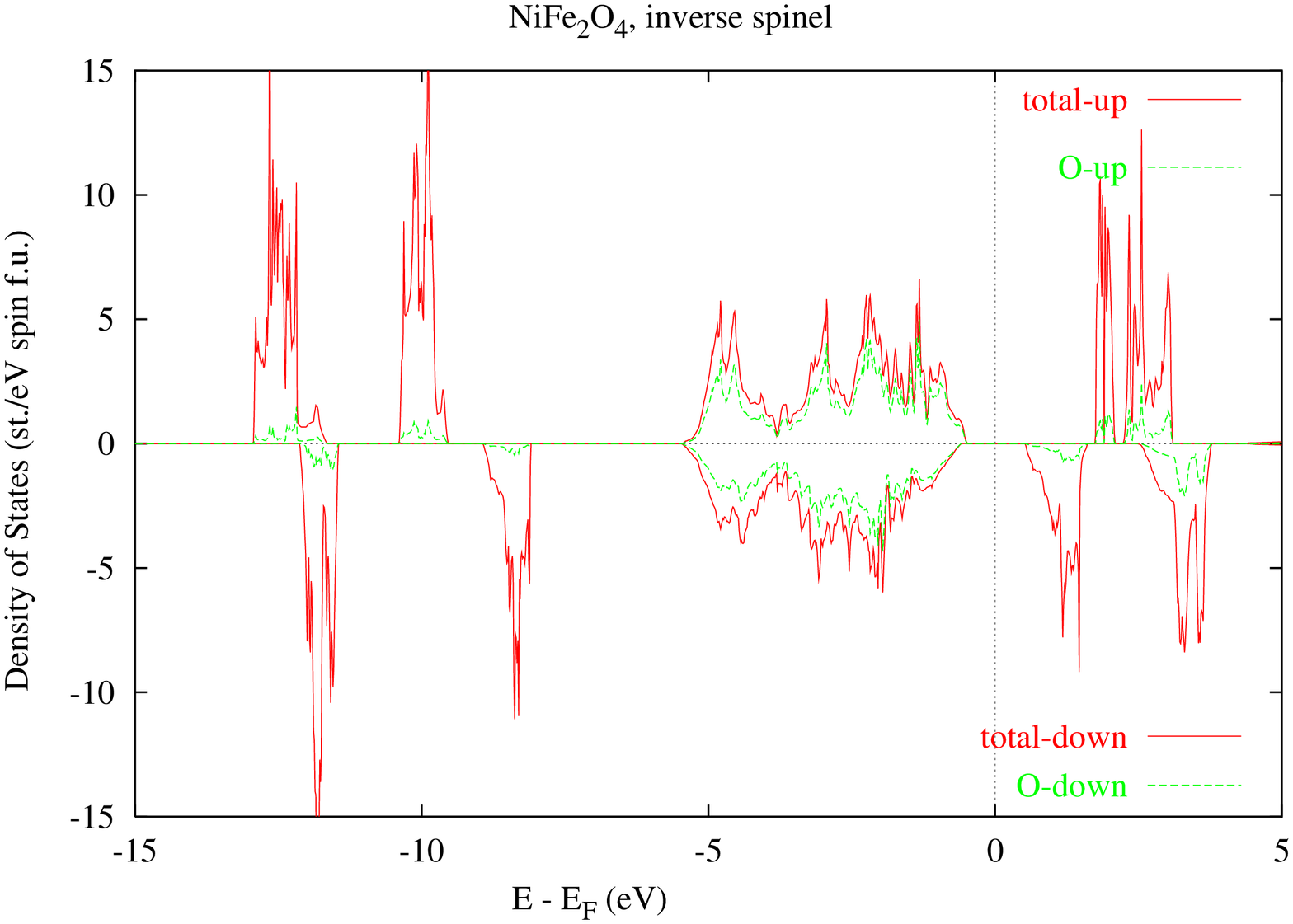}
\caption{(Color online) Spin decomposed total densities of states (in red), per formula unit, for 
NiFe$_{2}$O$_{4}$ in the normal spinel structure ('N') (top), NiFe$_{2}$O$_{4}$ in the inverse 
spinel structure for all 3+ ('I3+') scenario (middle), and NiFe$_{2}$O$_{4}$ in the inverse 
spinel structure ('I') (bottom). 
The oxygen contribution to the total density of states is also shown (green dotted lines).  
The minority DOS is printed on the negative side of the y-axis, while the majority contribution
is shown on the positive side of this axis.
\label{fig4}
}
\end{figure}

In agreement with experiments, we find the ground state of NiFe$_{2}$O$_{4}$ to
be insulating and of the inverse spinel kind (Table \ref{table2} and the bottom panel
of Fig. \ref{fig4}). The calculated energy gap is 0.98 eV (Table \ref{tableNi}).
It gets reduced to 0.26 eV in the normal spinel scenario, which is the most
energetically unfavourable solution for this compound. 
In the ground state 'I' scenario,
the octahedral B1-sites are occupied by Ni$^{2+}$  and B2 sites by Fe$^{3+}$
ions, while the tetrahedral sites are populated exclusively by Fe$^{3+}$ ions. Replacing 
the Co$^{2+}$ ions on the octahedral sites in CoFe$_{2}$O$_{4}$ by Ni$^{2+}$ leads to the reduction of 
the total spin magnetic moment
of 3 $\mu_{B}$ in CoFe$_{2}$O$_{4}$ to 2 $\mu_{B}$ in NiFe$_{2}$O$_{4}$ (Table \ref{table2}), 
since the spin magnetic
moment of the Ni$^{2+}$ ion is 1.57 $\mu_{B}$ (Table \ref{table6}), as compared to 
2.58 $\mu_{B}$ spin moment of the Co$^{2+}$ ion (Table \ref{table5}). The oxygen spin 
moments in both materials are comparable, and aligned in parallel to the cation spin 
moments on the octahedral sites. The width of the predominantly oxygen 2$p$ valence band 
in NiFe$_{2}$O$_{4}$ is comparable to the one of CoFe$_{2}$O$_{4}$, but is reduced with respect to the valence band
of magnetite. The reason being that the sixth localized $d$ electron of the Fe$^{2+}$ ion (in
Fig. \ref{fig2} (middle panel) seen at the bottom of the valence band between -5.0 and -6.0 
eV) is strongly hybridized with the 
oxygen $p$ band. The situation is different in NiFe$_{2}$O$_{4}$, where the three localized 
minority t$_{2g}$ electrons, seen at about -8.0 eV (Fig. \ref{fig4}, bottom panel), are 
well separated from the bottom of 
the valence band. Also, the exchange splitting of the conduction band, of importance to spin
filtering, is about 20\% smaller in NiFe$_{2}$O$_{4}$ than in the Verwey phase of
Fe$_{3}$O$_{4}$.

\begin{table}[h!]
\caption{Spin decomposed exchange splittings of the valence and conduction bands, as well
as the energy gaps (in eV), for both inverse and normal spinel structures for NiFe$_{2}$O$_{4}$.
Here VBM stands for the valence band maximum and CBM for the conduction band minimum, and
$\uparrow$ refers to spin-up- and $\downarrow$ to spin-down-component. 
}
\begin{tabular}{ccc}
\hline
 & 'I' scenario & 'N' scenario \\
\hline
VBM$^{\uparrow}$ - VBM$^{\downarrow}$    &  0.10  & -0.12 \\
CBM$^{\uparrow}$ - CBM$^{\downarrow}$    &  1.21  &  2.93  \\
CBM$^{\uparrow}$ - VBM$^{\uparrow}$      &  2.19  &  3.31  \\
CBM$^{\downarrow}$ - VBM$^{\downarrow}$  &  1.08  &  0.26  \\
Gap                                      &  0.98  &  0.26  \\
\hline
\end{tabular}
\label{tableNi}
\end{table}
 
\begin{table}[h!]
\caption{Type-decomposed spin magnetic moments (in $\mu_{B}$ per formula unit), 
calculated within SIC-LSD,
for NiFe$_{2}$O$_{4}$ for inverse and normal spinel scenarios. Here 'A' marks
the tetrahedral-, while 'B1' and 'B2' the octahedral-sites, and 'O1' and 'O2'
stand for two different types of oxygens.}

\begin{tabular}{cccccc}
\hline
Scenario & Fe$_{A}$$^{3+}$ & Ni$_{B1}$$^{2+}$ & Fe$_{B2}$$^{3+}$ & O1 & O2 \\
\hline
I    &  -4.11  &  1.57  &  4.11 &  0.14  & 0.07 \\
\hline
Scenario & Ni$_{A}$$^{2+}$ & Fe$_{B1}$$^{3+}$ & Fe$_{B2}$$^{3+}$ & O1 & O2 \\
\hline
N    &  -1.65  &  4.13  &  4.13 &  0.35 & 0.35 \\
\hline
\end{tabular}
\label{table6}
\end{table}

To understand details of the DOS of the 'I3+' scenario of NiFe$_{2}$O$_{4}$, it is helpful to follow 
the discussion of the same scenario for CoFe$_{2}$O$_{4}$, keeping in mind that a Ni$^{3+}$
ion has two minority $d$ electrons in addition to the five majority ones, localized by
SIC. Obviously, as already mentioned when discussing MnFe$_{2}$O$_{4}$, the positions of the localized
$d$ peaks calculated in the SIC-LSD should not be directly compared with photoemission
experiments.\cite{photoemission_comment}

The changes in the electronic structure of NiFe$_{2}$O$_{4}$ when moving from the
ground state inverse spinel structure to the normal spinel scenario are immediately
obvious from comparing the bottom and top panels of Fig. \ref{fig4}.
In the normal spinel case the Ni$^{2+}$ ions occupy the tetrahedral sites, while the 
octahedral sites are solely taken by the Fe$^{3+}$ ions. As a result, the total spin 
magnetic moment is increased from 2 $\mu_{B}$ per formula unit to 8 $\mu_{B}$ per 
formula unit, as seen in Table \ref{table1},
in agreement with experimental findings of Refs. \onlinecite{ulu,Manuel}.
However, as seen in the top panel of Fig. \ref{fig4}, the density of states is
still just insulating in both spin channels. 
Also, unlike in the case of the inverse spinel structure, the valence band is strongly 
spin-polarized, and the polarization is negative. The oxygen spin magnetic moment is 
0.35 $\mu_{B}$, aligned in parallel to the 
Fe spin moment (see Table \ref{table6}), and three times the value it has in the inverse 
spinel structure. Also, the
Ni spin moment is slightly increased in magnitude to -1.65 $\mu_{B}$. Moreover, the 
exchange splitting of the conduction band is more than twice increased in the normal spinel, 
in comparison with the inverse spinel structure, from 1.21 eV to 2.93 eV. As in the case
of the other spinel ferrites in
realizing Ni$^{2+}$ ions, the e$_{g}$ states are populated first, i. e., are lying lower in 
energy than the t$_{2g}$ states, which is opposite to the inverse spinel structure. 
However, energetically, the normal spinel structure for NiFe$_{2}$O$_{4}$ is very unfavourable
with respect to the inverse spinel structure (Table \ref{table1}). 

Including the spin-orbit coupling for the ground state inverse spinel scenario gives
rise to the total orbital moment of 0.67 $\mu_{B}$ per formula unit, and is mostly
due to Ni ions, with some minor contributions from Fe atoms. This calculated value is
over two times larger than calculated from LSD in Ref. \onlinecite{guo2}. 
Also, in the earlier SIC-LSD calculations for TM oxides Svane and Gunnarsson~\cite{ASOG} 
obtained the orbital moment for NiO of 0.27 $\mu_{B}$, which is substantially smaller 
than in the present calculations. Since the total spin moment is slightly reduced to 
1.9997 $\mu_{B}$ per formula unit, when SOC is taken into account, hence the ratio of 
the total orbital and spin moments is calculated to be about 0.34. 
The latter is in good agreement with the experimental 
estimates of 0.27$\pm$0.07~\cite{gerrit} and 0.34 (within error bars of up to 
$\pm$0.11) for Ni in NiFe$_{2}$O$_{4}$ and NiO.~\cite{fernandez,gerrit}
Note that the orbital moment due to the Ni$^{2+}$ ion alone is about 0.7 $\mu_{B}$, while 
its spin moment is 1.58 $\mu_{B}$, both per formula unit, giving rise to the orbital to
spin moment ratio of 0.44 for this ion. The total orbital moment is mostly due to the 
Ni$^{2+}$ ion, as the contributions of other TM ions are smaller by an order of magnitude 
or so. \\

%

\section{Conclusions}

We have shown that, owing to a better treatment of correlations,
SIC-LSD can provide useful insights to the nature of a number of spinel ferromagnetic
insulators. We have been able to address the issues of the normal versus inverse
spinel arrangements in these systems, their electronic and magnetic properties 
and the valence of the transition metal atoms. 
We find all the studied ferrites to be insulating for both the inverse and normal 
spinel scenarios, with the calculated energy gaps being smaller in the normal 
spinel environment, however showing an increasing 
trend when moving from MnFe$_{2}$O$_{4}$ to NiFe$_{2}$O$_{4}$. We have observed dramatic increase of the 
calculated spin magnetic moments, as well as 
the exchange splitting of the conduction bands, when moving from the inverse to 
normal spinel scenarios, some of which have been observed in experiments.~\cite{Manuel}

The total energy considerations
seem to favour the inverse spinel arrangement of TM ions as the ground state
configurations for all the studied ferrites, with the possible exception of MnFe$_{2}$O$_{4}$
where the normal spinel environment may be preferred, with Mn$^{2+}$ ions on
the tetrahedral sites and the Fe$^{3+}$ ions on the octahedral sublattice.
Also, based on the total energy arguments, we find a partial delocalization 
of the minority spin states to be favourable in Fe$_{3}$O$_{4}$ and CoFe$_{2}$O$_{4}$, leading to the half-metallic
ground states with all TM ions in the trivalent configuration and in the inverse spinel
arrangement. The fully inverse spinel scenario, with the Fe$^{3+}$ ions on the 
tetrahedral sites and the octahedral sites occupied both by Ni$^{2+}$ ions and
Fe$^{3+}$ ions, is found to be the ground state only in NiFe$_{2}$O$_{4}$. Finally, these findings 
constitute a good starting point for further studies, incorporating alloying of
the normal and inverse spinel structures.


\section*{Acknowledgements}
LP was supported by the Division of Materials Science an Engineering, Office of 
Basic Energy Sciences, US Department of Energy.


\begin{thebibliography}{99}
\bibitem{photoemission_comment} To improve the agreement between the calculated position of the
TM $d$ states from the SIC-LSD method and the photoemission measurement, one can perform
$\Delta_{SCF}$ calculations, that is calculations of the total energy difference between two 
configurations of TM assumed to represent well the initial and final states of a photoemission 
experiment. This is expressed by the formula
%
$\Delta_{SCF} = E(d^{n-1}, N_{val} + 1) - E(d^{n}, N_{val})$,
%
meaning that one removes an electron from a TM $d$ level and introduces it at the top of the valence
band, thus increasing the total number of valence electrons, $N_{val}$, by 1. Here $E$ represents the
total energies of the respective $d^{n-1}$ and $d^{n}$ configurations.
Of course, a problem one faces with this kind of estimate is that the correct final state
might not be the ground state of the system with one electron removed. \\
Another way of improving the agreement with spectroscopies is to apply the OEP philosophy.
The SIC-LSD method is an orbital dependent density functional theory. Formally, however, 
SIC-LSD may be viewed as a standard density functional theory, implying that the SIC-LSD
energy functional can be represented as a functional of the total charge density alone
and minimized with respect to it.
This means that there exists an effective Kohn-Sham equation with an effective potential,
which is common to all Kohn-Sham states, is self-interaction free and
depends only on the total charge density. Here the situation is completely analogous to the
optimized effective potential introduced in connection with the Hatre-Fock
approximation,\cite{iafrate,kotani} for which case the Kohn-Sham eigenvalues are often
compared to quasiparticle energies, with a considerable improvement over the Hartree-Fock
eigenenergies.
Adopting the OEP philosophy, one can search for the effective potential, which reproduces
the SIC spin-density, and with such a potential derive the density of the localized TM $d$ states.
The search may be constrained by looking only for that particular potential shift on the TM sites, 
which will reproduce the self-consistent TM spin moment of the SIC-LSD calculations.
\bibitem{gap1} Note that our SIC-LSD calculations for the Verwey scenario and the true low
temperature monoclinic phase obtained a gap of $\sim$0.1 eV,\cite{Szotek1} which compares 
well with the experimental value of 0.14 eV~\cite{gap}.
\end{thebibliography}
\end{document}